\documentclass[superscriptaddress,showpacs,twocolumn,nofootinbib,aps]{revtex4-1}
\usepackage[normalem]{ulem}
\usepackage{graphicx}
\usepackage{color}
\usepackage{xcolor}
\usepackage{verbatim}
\usepackage{subfigure}
\usepackage{amssymb}
\usepackage{amsmath}
\usepackage{comment}
\usepackage{chngcntr}
\usepackage{appendix}
\usepackage{lipsum}
\usepackage{dsfont}
\usepackage{xfrac}
\usepackage{xr}
\usepackage[makeroom]{cancel}
\newcommand{\bxi}{\boldsymbol{\xi}}

\newcommand{\blambda}{\boldsymbol{\lambda}}
\newcommand{\beeta}{\boldsymbol{\eta}}

\newcommand{\bzeta}{\boldsymbol{\zeta}}
\newcommand{\dd}{\text{d}}

\newcommand{\ee}{\text{e}}

\newcommand{\br}{{\bm{r}}}

\newcommand{\bu}{{\bm{u}}}
\newcommand{\bR}{{\bm{R}}}

\newcommand{\bnabla}{\boldsymbol{\nabla}}
\newcommand{\bmu}{\boldsymbol{\mu}}

\newcommand{\id}{\mathbf{1}}

\newcommand{\bF}{{\bm{F}}}

\newcommand{\bj}{{\bm{j}}}

\newcommand{\bA}{\text{\bf A}}

\newcommand{\bP}{{\bm{P}}}

\newcommand{\Teff}{T^{\text{eff}}}

\newcommand{\Gammaavg}{\langle\Gamma\rangle_{\text{b}}}
\newcommand{\bFavg}{\langle\bF\rangle_{\text{b}}}
\newcommand{\rhoB}{\rho_\text{b}}
\newcommand{\rhoo}{\rho_\text{o}}
\newcommand{\rhooss}{\rho_{\text{o}}^\text{ss}}

\newcommand{\subfigref}[2]{\hyperref[#1]{\ref*{#1}#2}} 

\usepackage{amsthm}
\usepackage{bm}
\usepackage[hang,flushmargin]{footmisc}
\usepackage[pdftex]{hyperref}
\begin{document}

\title{Odd dynamics of passive objects in a chiral active bath}
\author{Cory Hargus}
\affiliation{Laboratoire Mati\`ere et Syst\`emes Complexes (MSC), Université Paris Cité  \& CNRS (UMR 7057), 75013 Paris, France}
\author{Federico Ghimenti}
\affiliation{Laboratoire Mati\`ere et Syst\`emes Complexes (MSC), Université Paris Cité  \& CNRS (UMR 7057), 75013 Paris, France}
\affiliation{Department of Applied Physics, Stanford University, Stanford, CA 94305, USA}
\author{Julien Tailleur}
\affiliation{Laboratoire Mati\`ere et Syst\`emes Complexes (MSC), Université Paris Cité  \& CNRS (UMR 7057), 75013 Paris, France}
\affiliation{Department of Physics, Massachusetts Institute of Technology, Cambridge, MA 02139, USA}
\author{Frédéric van Wijland}
\affiliation{Laboratoire Mati\`ere et Syst\`emes Complexes (MSC), Université Paris Cité  \& CNRS (UMR 7057), 75013 Paris, France}
\affiliation{Yukawa Institute for Theoretical Physics, Kyoto University,
Kitashirakawaoiwake-cho, Sakyo-ku, Kyoto 606-8502, Japan}
\begin{abstract}
When submerged in a chiral active bath, a passive object becomes a spinning ratchet imbued with odd transport properties. We present the most general Langevin dynamics for a rigid body in a chiral active bath, in the adiabatic  limit of large object mass. For rotationally symmetric objects, odd diffusion and odd mobility are connected by an Einstein relation, that we show numerically to break down outside the adiabatic limit. As the object symmetry decreases, its dynamics becomes increasingly irreversible: a massive disk exhibits an effective equilibrium dynamics, while a rod admits distinct translational and rotational temperatures, and a wedge is fully irreversible.
Conversely, this departure from equilibrium can be read in universal far-field currents and density modulations of the bath, which we measure numerically and derive analytically.
\end{abstract}

\maketitle

\vspace{0.1in}
A microscopic object in a fluid bath inherits its dynamics from collisions with the bath particles.
When the bath is in equilibrium, the object obeys the Einstein relation and exhibits Boltzmann statistics.
Active baths, by contrast, allow a richer set of phenomena and have thus attracted a lot of attention~\cite{wu,loi2008effective,Underhill2008,Leptos2009,Dunkel2010,Kurtuldu2011,Foffano2012,Mino2013,Morozov2014,Burkholder2017,Chaki2018,Kanazawa2020,Knezevic2020,Reichert2021,maes,Granek2022, Solon2022,sorkinSecondLawThermodynamics2024}.
Notable phenomena include ratchet motions~\cite{magnasco1993forced,van2004microscopic,angelani2011active,gnoli2013brownian,reichhardt2017ratchet}, demonstrating how an active bath can power microscopic
engines~\cite{krishnamurthy2016micrometre,zakine2017stochastic,pietzonkaAutonomousEnginesDriven2019,fodorActiveEnginesThermodynamics2021}. When asymmetric gears are inserted into an active bacterial bath~\cite{di2010bacterial,sokolov2010swimming,anandTransportEnergeticsBacterial2024a}, rotational ratchet motion arises due to broken time-reversal symmetry of the bath and broken parity symmetry ({\it i.e.} chirality) of the object.

More recent investigations have looked at chiral active baths, in which both of these symmetries are broken by the bath itself.
In such baths, even symmetric objects become 
rotational ratchets~\cite{li2023chirality} imbued with
odd diffusivity~\cite{Hargus2021,Kalz2022} and odd mobility~\cite{Reichhardt2019,Poggioli2023odd}.
While extensive analytical and numerical results are available for achiral baths, little is known theoretically of the dynamics of passive objects embedded in chiral active fluids. The latter are however abundant in both biological~\cite{drescher2009dancing,petroff2015fast,beppu2021edge,tan2022odd} and synthetic active systems~\cite{Soni2019,bililign2022motile}.
This raises two key questions: how are the emerging properties of the object (and the bath) interrelated, and how are they influenced by the object shape?

\begin{figure}
    \includegraphics[width=.49\textwidth]{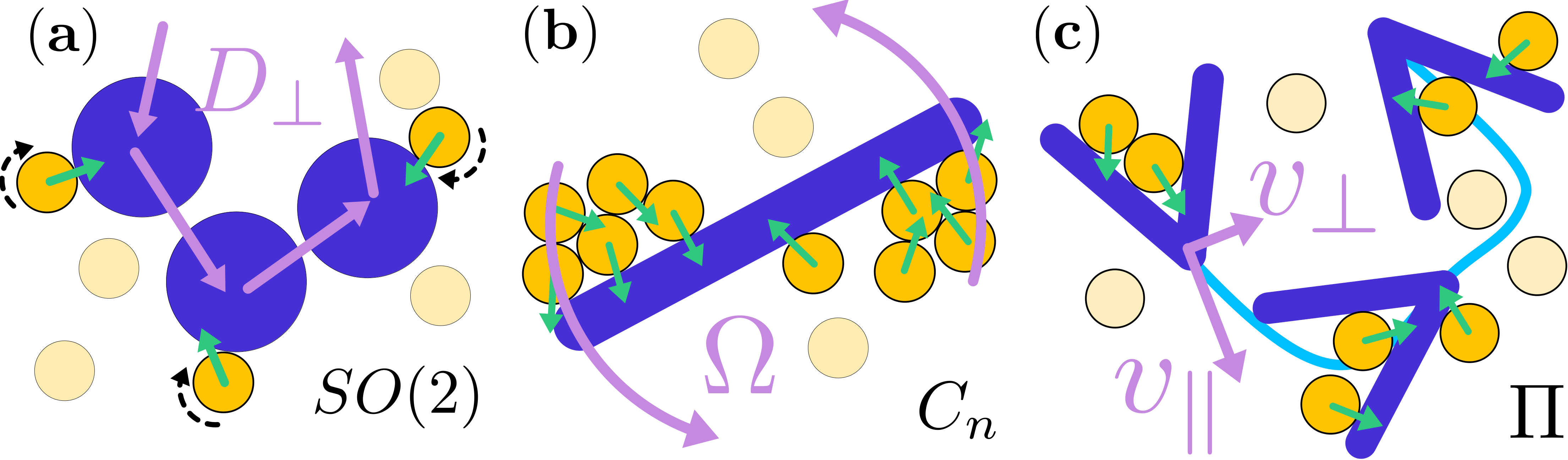}
    \caption{\textbf{Passive objects}
    inherit chiral dynamics from the bath. A disk \textbf{(a)} exhibits odd diffusivity $D_\perp$, a rod \textbf{(b)} additionally rotates with a ratchet velocity $\Omega$, and a passive wedge \textbf{(c)} additionally translates with ratchet velocities $v_\parallel$ and $v_\perp$.
    \label{fig:cartoon}}
    \end{figure}

In this Letter, we answer these questions by constructing a general theory to describe the dynamics of a massive object embedded in a chiral active bath. This allows us to show that the object dynamics become increasingly rich as its symmetry decreases. Specifically, we consider the disk, rod, and wedge depicted in Fig.~\ref{fig:cartoon}. For the disk, we find an effective equilibrium description for the translational degrees of freedom, accompanied by circulating currents in momentum space.
The rod departs from this effective equilibrium, by displaying distinct temperatures for the rotational and translational degrees of freedom.
Last, the wedge is fully irreversible, behaving as both a translational and rotational ratchet.
Finally, we show how the object symmetry influences the bath itself through the generation of long-range density modulation and currents.
Here we focus on the presentation of our results; numerical details are provided in the End Matter while detailed derivations can be found in a companion paper~\cite{companionPRE}.

\vspace{0.1in}
\noindent\textbf{\textit{Model system and effective dynamics.}}
We consider an extended rigid body of arbitrary shape interacting under Hamiltonian dynamics with a chiral active bath in two dimensions. This object has mass $M$ and moment of inertia $I$. Its state is characterized by its position $\bR$, momentum $\bP$, orientation $\Theta$ and angular momentum $L$. The equations of motion of the object are
\begin{equation}\label{eq:EOM-newtonian}
M \ddot \bR = \dot \bP = \bF\,,\quad I \ddot \Theta =\dot L = \Gamma\,,
\if{	\begin{split}
		\dot \bR &= \frac{1}{M}\bP\,, \\
        \dot \bP &=  \bF\,, \\
     \end{split}
     \quad \quad
     \begin{split}
        \dot \Theta &= \frac{1}{I}L\,, \\
		\dot L &= \Gamma\,,
	\end{split}}\fi
\end{equation}
where $\bF$ and $\Gamma$ are the total force and torque due to
the bath particles.  The latter are modeled as
chiral active Brownian particles, whose positions $\bm{r}_i$ and
orientations $\theta_i$ evolve as
\begin{equation}\label{eq:cABP-EOM}
        \gamma \dot\br_i = \bF_{i} + f_0 \bu_i\,,\quad
        \dot{\theta}_i = \omega_0 + \sqrt{2 D_r} \xi_i\,.
\if{
    \begin{split}
        \gamma \dot\br_i &= \bF_{i} + f_0 \bu_i + \sqrt{2 \gamma D_t} \beeta_i\,,\\
        \dot{\theta}_i &= \omega_0 + \sqrt{2 D_r} \xi_i\,.
    \end{split}}\fi
\end{equation}
Here, $\gamma$ is the substrate friction,  $f_0$ is the active force
oriented along $\bu_i = [\cos(\theta_i),
  \sin(\theta_i)]^T$, $\omega_0$ is an angular drift, $D_r$ is the rotational diffusivity, and $\xi_i$ are centered unit-variance Gaussian white noises.
The bath displays two active lengthscales: the persistence length $\ell_p = \frac{f_0}{D_r \gamma}$ and gyroradius $\ell_g = \frac{f_0}{|\omega_0| \gamma}$. Object-bath interactions are reciprocal, with
$\sum_i \bF_{i} = -\bF$.

If the motion of the object is very slow relative to that of the bath,
we can work in the adiabatic
limit~\cite{van1986brownian,granek2020,Solon2022,jayaram2023effective}
where the probability distribution of the full system factorizes as
$\rho(\bR,\bP,\Theta,L,\br^N,\bu^N,t) \approx
\rhoo(\bR,\bP,\Theta,L,t) \rhoB(\br^N,\bu^N|\bR, \Theta)$. This leaves
all time dependence in $\rhoo$, while the bath
relaxes to the instantaneous object
configuration. This separation of timescales is characterized by the
dimensionless parameter $\epsilon = \sqrt{\gamma/D_r M}$, which is
small in the limit of massive objects. Following established
procedures~\cite{van1986brownian,Solon2022}, we integrate out the bath
degrees of freedom to obtain underdamped Langevin dynamics for the
object:
\begin{equation}\label{eq:tracerdynamics}
  \begin{split}
    \begin{bmatrix} \dot\bP \\ \dot L \end{bmatrix} &= \begin{bmatrix} \bFavg \\ \Gammaavg\end{bmatrix} - \underbrace{\begin{bmatrix} \bzeta_{\bP\bP} & \bzeta_{\bP L} \\ \bzeta_{L\bP} & \zeta_{LL} \end{bmatrix} }_{\bzeta}\begin{bmatrix} \frac{1}{M}\bP \\ \frac{1}{I} L \end{bmatrix} +\underbrace{\begin{bmatrix}\bxi_P(t) \\ \xi_L(t)\end{bmatrix}}_{\bxi}\,.
  \end{split}
\end{equation}
The bath produces three effects on the object: a mean force $\bFavg$ and torque $\Gammaavg$ (where $\langle \cdot \rangle_{\text{b}}$ denotes an average at fixed object configuration), an instantaneous friction $\bzeta$, and Gaussian noise $\bxi$. In this framework, the  nine coefficients entering the friction matrix $\bzeta$ are given by an Agarwal formula~\cite{agarwal1972}
\begin{equation}\label{eq:kubo-agarwal}
    \bzeta \hspace{-1mm} = \hspace{-1.5mm} \int_0^{\infty}\hspace{-3.5mm}\dd t\begin{bmatrix} \langle \delta\bF(t) \bnabla_\bR \ln \rho_\text{b} (0)\rangle_\text{b} & \langle \delta\bF(t)\partial_\Theta\ln \rho_\text{b} (0)\rangle_\text{b}\\ \langle \delta\Gamma(t)\bnabla_\bR^T\ln \rho_\text{b} (0)\rangle_\text{b} & \langle \delta\Gamma(t)\partial_\Theta\ln \rho_\text{b} (0)\rangle_\text{b} \end{bmatrix}\hspace{-1mm},
\end{equation}
where $\delta\bF=\bF-\bFavg$ and $\delta\Gamma=\Gamma-\Gammaavg$.
The Gaussian noises are characterized by their correlation matrix $\blambda$, defined by
\begin{equation}\label{eq:noise}
    \langle \bxi(t) \bxi(t') \rangle = \blambda \delta_+(t-t') + \blambda^T\delta_-(t-t')\,,
\end{equation}
where $\delta_\pm$ is the restriction of the Dirac function to
$\mathbb{R}_\pm$~\cite{chun2018emergence} and $\blambda$ satisfies the Green-Kubo
formula
\begin{equation}\label{eq:green-kubo}
    \blambda = \int_0^{\infty}\dd t \begin{bmatrix} \langle \delta\bF(t)\delta\bF(0)\rangle_\text{b} & \langle\delta\bF(t) \delta\Gamma(0)\rangle_\text{b} \\ \langle \delta \Gamma(t) \delta\bF^T(0)\rangle_\text{b} & \langle\delta\Gamma(t) \delta\Gamma(0)\rangle_\text{b}\end{bmatrix}\,.
\end{equation}

\begin{figure}
    \centering
    \includegraphics[width=.49\textwidth]{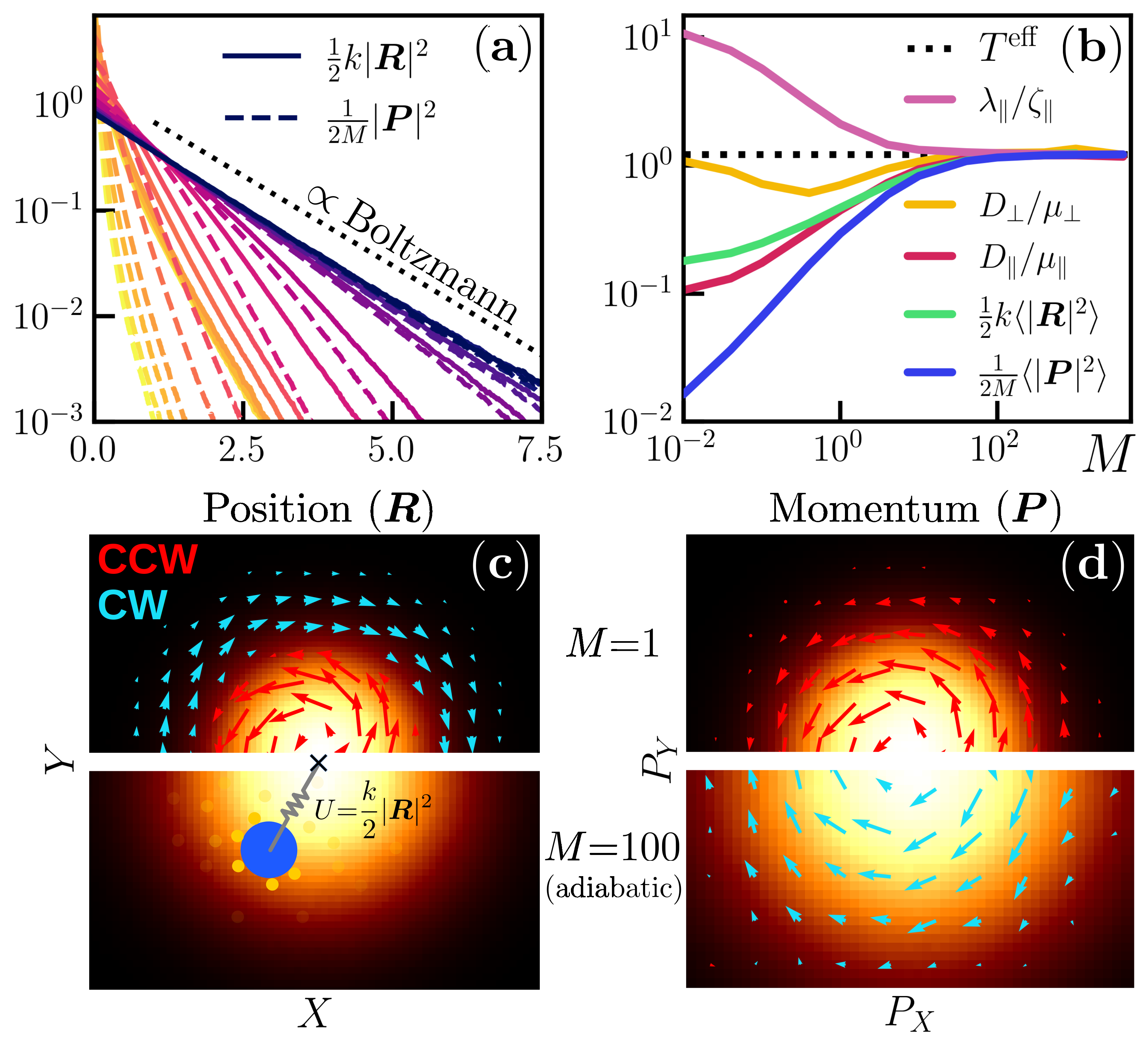}
    \caption{\textbf{Confined disk: effective equilibrium and circulating currents} ($\ell_p=10$, $\ell_g=5$).
      \textbf{(a)} The disk's potential (solid) and kinetic (dashed)
      energies converge to the Boltzmann distribution in the adiabatic
      limit ($M \in [10^{-2}, 10^{3}]$ increasing from light to dark).
      \textbf{(b)} Even and odd thermometers, constructed from
      $\bm{\lambda}$, $\bm{\zeta}$, $\mathbf{D}$, and $\bm{\mu}$, all
      converge to a single $\Teff$ in the adiabatic limit.
      \textbf{(c)} Probability density $\rho_\bR$ (heatmap) and flux
      $\bm{J}_\bR^{\text{ss}}$ (arrows) of the confined disk (cartoon
      overlay).  Steady circulating currents (top) vanish at large $M$
      (bottom) due to the odd Einstein relation~\eqref{eq:einstein}.
      \textbf{(d)} $\rho_\bP$ and $\bm{J}_\bP^{\text{ss}}$ of the same
      system. Circulation persists even in the adiabatic limit. The
      direction of rotation, determined by competition between
      $\zeta_\perp$ and $\lambda_\perp$, reverses at smaller $M$.
    \label{fig:disk-adiabaticity}
    }
\end{figure}

Equations~\eqref{eq:tracerdynamics}-\eqref{eq:green-kubo} provide
the most general dynamics for a rigid body in a chiral active bath,
within the adiabatic approximation. They hold in any dimension $d\geq
2$ but neglect long-time tails due to conservation
laws~\cite{VanBeijeren1982,dorfman2021contemporary}, that can play an
important role in low-dimensional active
systems~\cite{granek2020}. The object dynamics is then entirely
determined by $\bzeta$, $\blambda$, $\bFavg$ and $\Gammaavg$, which are set by the
object shape. 
We now apply these theoretical results to show how the object departs from an effective equilibrium limit as its symmetry decreases.

\vspace{0.1in}
\noindent\textbf{\textit{The $SO(2)$ swirling disk}.}  We begin by
considering a circular disk, which has only translational degrees of
freedom. By symmetry, $\bFavg={\bf 0}$, and Eq.~\eqref{eq:tracerdynamics} reduces to
\begin{equation}\label{eq:disk-langevin}
    \dot\bR = M^{-1} \bP\,, \hspace{4mm} \dot\bP = -M^{-1}\bzeta_{\bP\bP} \bP + \bxi_\bP(t)\,.
\end{equation}
The combined isotropy of the bath and the disk make $\bzeta$ and $\blambda$ belong to the space of isotropic $2\times 2$ matrices, which is spanned by $\id=\begin{bmatrix} 1 &0\\0&1\end{bmatrix}$ and $\bA=\begin{bmatrix} 0 &-1\\1&0\end{bmatrix}$, so that $\bzeta_{\bP\bP} = \zeta_\parallel \id + \zeta_\perp \bA$ and $\blambda_{\bP\bP} = \lambda_\parallel \id + \lambda_\perp \bA$. Here, $\zeta_\parallel$ and $\lambda_\parallel$ are the usual friction and noise correlations, respectively. Their odd counterparts, $\zeta_\perp$ and $\lambda_\perp$, stem from the breaking of time-reversal and parity symmetry of the bath~\cite{hargusFluxHypothesisOdd2025}. As we show below, these drive steady circulating currents of the object, providing a counterpoint to those observed in confined chiral active fluids~\cite{vanzuidenSpatiotemporalOrderEmergent2016,souslovTopologicalSoundActiveliquid2017,dasbiswasTopologicalLocalizationOutofequilibrium2018,Soni2019,Yang2021,Vuijk2020,kalz2024oscillatory,VegaReyes2022,Caprini2019,capriniChiralActiveMatter2023,liRobustEdgeFlows2024,wittmannConfinedActiveParticles2024}.

To investigate these odd currents, we confine the disk using a harmonic potential $U(\bR) = \frac{1}{2}k|\bR|^2$, as shown in Fig.~\ref{fig:disk-adiabaticity}. The simulations in Fig.~\subfigref{fig:disk-adiabaticity}{a} show how the nonequilibrium dynamics of light disks are replaced at large $M$ by an effective equilibrium one, leading to the Boltzmann steady-state solution of Eq.~\eqref{eq:disk-langevin} given by
\begin{equation}\label{eq:boltzmann}
    \rhooss = \rho_\bR(\bR) \rho_\bP(\bP) \propto \ee^{-\left(U(\bR) + \frac{1}{2M}|\bP|^2 \right) / \Teff}\,,
\end{equation}
where $\Teff=\frac{1}{2M}\langle |\bP|^2 \rangle=\frac{1}{2}k\langle |\bR|^2 \rangle$ is an effective temperature.

The flux in $\bR$-space is $\bm{J}_\bR^{\text{ss}} = \left(\Teff \bmu
- \mathbf{D}\right)\bnabla_\bR \rho_\bR$, where $\mathbf{D} =
D_\parallel \id + D_\perp \bA$ is the diffusivity computed
from the Green-Kubo relation $\mathbf{D} = \frac{1}{M^2}\int_0^\infty
dt\ \langle \bP(t) \bP(0) \rangle$.  The odd diffusivity $D_\perp$
generates a flux of $\rho_\bR$ perpendicular to its
gradient~\cite{Hargus2021}. Conversely, the mobility $\bmu =
\mu_\parallel \id + \mu_\perp \bA = \bzeta^{-1}$ predicts the
drift under an external force $\bm{F}^{\text{ext}}$ as $\langle \bP
\rangle_{\bm{F}^{\text{ext}}} = M \bmu \bm{F}^{\text{ext}}$, where the
odd mobility $\mu_\perp$ causes drift perpendicular to
$\bm{F}^{\text{ext}}$~\cite{Reichhardt2019,Poggioli2023odd}, and is
maximal when the disk diameter is of the order $\ell_g$ (see Fig.~\ref{fig-EM:disk-end-matter} below). Einstein relations then follow
as
\begin{equation}\label{eq:einstein}
    D_\parallel = \Teff\mu_\parallel\,, \quad \quad D_\perp = \Teff\mu_\perp\,,
\end{equation}
remarkably holding for the odd parts as well as the even. Thus, while
$D_\perp$ and $\mu_\perp$ can individually drive currents satisfying
$\bnabla_\bR \cdot \bm{J}_\bR^{\text{ss}} = 0$, these exactly cancel
at large $M$, as shown in Fig.~\subfigref{fig:disk-adiabaticity}{c}
(bottom panel), leading to an effective equilibrium reminiscent of that observed for a bath of achiral active particles~\cite{Solon2022}. In contrast, when $M$ is small, the disk departs from the equilibrium distribution
of Eq.~\eqref{eq:boltzmann}. Competition between $D_\perp$ and
$\mu_\perp$ then results in steady flows illustrated in
Fig.~\subfigref{fig:disk-adiabaticity}{c} (top panel).

\begin{figure}[t]
    \centering
    \includegraphics[width=.49\textwidth]{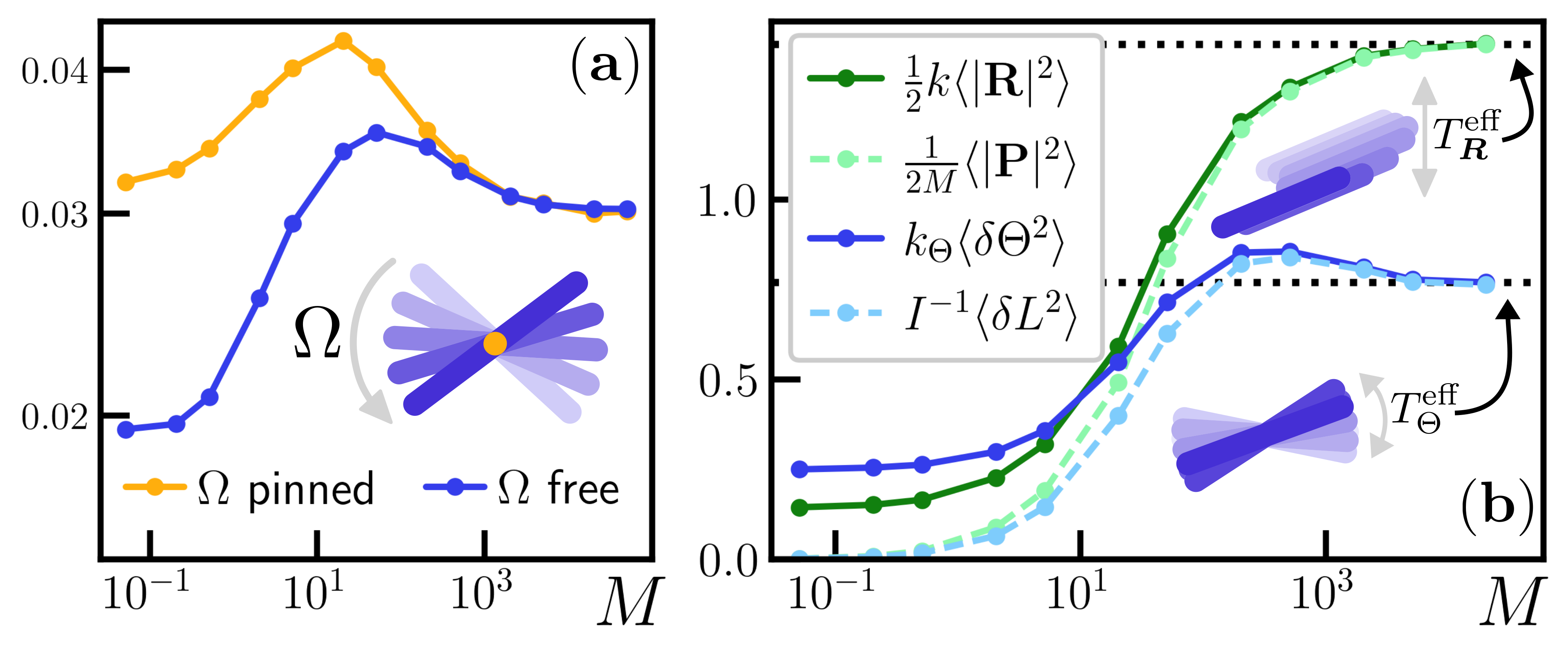}
    \caption{\textbf{Ratchet motion and effective temperatures of the spinning rod} ($\ell_\mathrm{rod}=5$,
      $\ell_p=10$, and $\ell_g = 5$). \textbf{(a)} The angular velocities of
      rods pinned at the center or free to translate
      agree at large $M$. \textbf{(b)} A rod confined to a
      harmonic potential exhibits rotational (blue) and translational
      (green) dynamics that independently recover equilibrium-like
      statistics at large $M$, with the mean potential (solid) and
      kinetic (dashed) energies agreeing, but at two distinct
      effective temperatures $\Teff_\bR$ and $\Teff_\Theta$.
    \label{fig:rod-rotation}
    }
\end{figure}

In the adiabatic limit, the flux of $\rho_\bP$ is $\bm{J}^{\text{ss}}_\bP = \left(\Teff\bzeta - \blambda\right) \bnabla_\bP \rho_\bP$.  The continuity condition $\bnabla_\bP \cdot \bm{J}^{\text{ss}}_\bP = 0$ requires that the even parts of $\blambda$ and $\bzeta$ be related by a second fluctuation-dissipation theorem, but places no such requirement on the odd parts:
\begin{equation}\label{eq:2FDT}
\lambda_\parallel = \Teff\zeta_\parallel\,, \quad \quad \lambda_\perp \ne \Teff\zeta_\perp\,.
\end{equation}
The inequality in Eq.~\eqref{eq:2FDT} results in odd currents that persist even in the adiabatic limit (Fig.~\subfigref{fig:disk-adiabaticity}{d}). This mismatch is a unique odd signature of chiral nonequilibrium baths.

We stress that Eqs~\eqref{eq:boltzmann}-\eqref{eq:2FDT} provide
independent thermometers to measure the effective temperature of the
system (Fig.~\subfigref{fig:disk-adiabaticity}{b}). Beyond the adiabatic regime, all involved quantities remain
well-defined and are of rheological
interest~\cite{saintillanRheologyActiveFluids2018,sheaForceRenormalizationProbes2024},
but cease to be directly connected through an effective temperature.

\begin{figure}
    \centering
    \includegraphics[width=.5\textwidth]{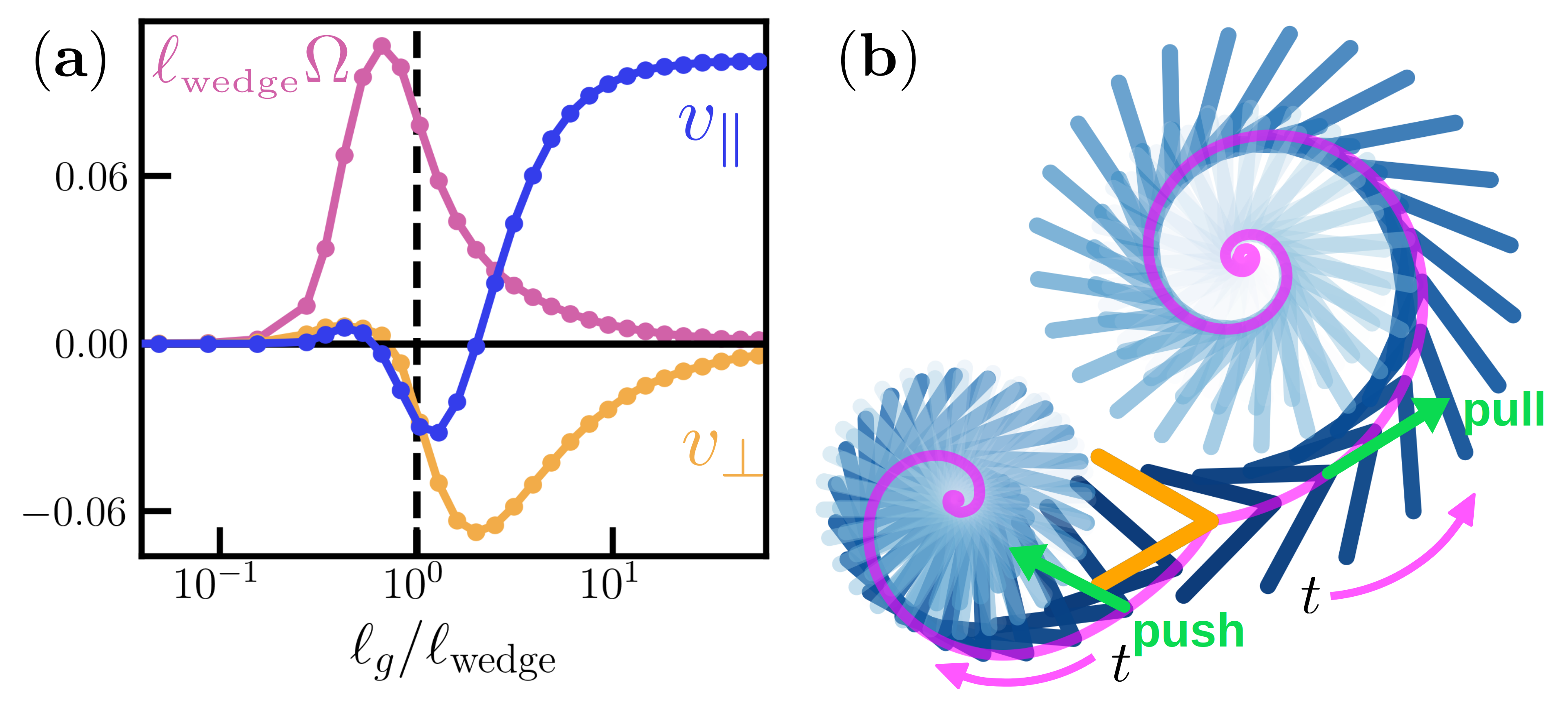}
    \caption{\textbf{Nonequilibrium wedge dynamics} ($\ell_p=10$, $\ell_\mathrm{wedge} = 3.6$, $M=100$).
    \textbf{(a)} Ratchet speeds $|v_\perp|$ and $|\Omega|$ are maximized at $\ell_g \approx \ell_\mathrm{wedge}$.
    \textbf{(b)} Average trajectories with $\ell_\mathrm{wedge}=\ell_g$, conditioned on the initial position (orange) when pushed or pulled with $\bm{F}^\mathrm{ext} = \pm 0.1 \bm{u}(\Theta)$  (green arrows). Pink lines trace the tip path. Asymmetry of the two trajectories results from rotation-translation coupling in $\bzeta$.
    \label{fig:wedge}
    }
\end{figure}

\begin{figure*}[t]
    \centering
    \includegraphics[width=\textwidth]{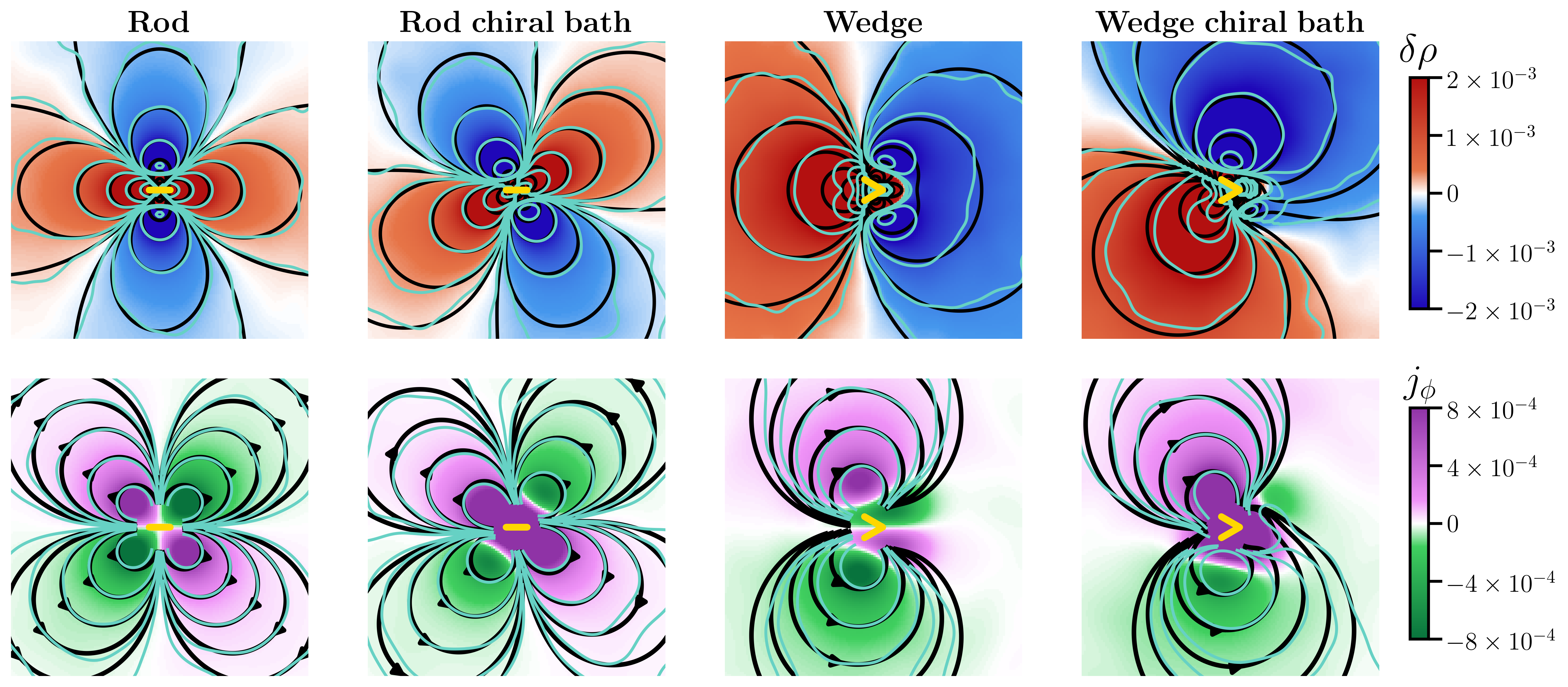}
    \caption{\textbf{Far-field density and currents}.
    Top row: Steady-state density (heatmap and blue contour lines) compared against the multipole prediction (black contour lines).
    Bottom row: Steady-state current (blue streamlines) and heatmap of the rotational component $j_\phi = \bm{r}\times\bj/|\bm{r}|$ compared against multipole prediction (black streamlines).
    First and third columns: achiral ($\ell_p=10$, $\ell_g = \infty$). Second and fourth columns: chiral ($\ell_p=10$, $\ell_g = 10$).
    }
    \label{fig:multipole}
\end{figure*}

\vspace{0.1in}
\noindent\textbf{\textit{The $C_2$ spinning rod}.}  We now proceed beyond the isotropic disk to consider the rod pictured in Fig.~\subfigref{fig:cartoon}{b}. 
The chiral bath now drives asymmetric accumulation around the object, leading to a net torque. The full system, however, maintains two-fold rotational symmetry ($C_2$). Consequently, $\bFavg = 0$, a result that holds for any object with $n$-fold rotational symmetry ($C_n$).

Furthermore, Eqs.~\eqref{eq:kubo-agarwal} and~\eqref{eq:green-kubo} show
the elements of $\bzeta$ and $\blambda$ that couple rotational and
translational motion to vanish under $C_n$ symmetry. This decoupling
means that the translational dynamics of the rod are structurally
identical to those of the isotropic disk, while the rotational
dynamics obeys
\begin{equation}\label{eq:rod-langevin}
    \dot L= \Gammaavg - I^{-1}\zeta_{LL}L + \xi_L(t)\,,
\end{equation}
leading to a non-zero average angular velocity $\Omega = \Gammaavg/\zeta_{LL}$. This unexpected independence between translational and rotational dynamics is demonstrated in Fig.~\subfigref{fig:rod-rotation}{a}, where a rod pinned at its center is shown to spin at the same speed as an unpinned rod (only) in the adiabatic limit. The freely rotating rod is then Boltzmann-distributed as $\rhooss \propto e^{-\delta L^2/2I\Teff_\Theta}$, where $\delta L=L - I\Omega$ and $\Teff_\Theta = I^{-1} \langle \delta L^2\rangle$ is the rotational temperature.

Another consequence of this decoupling is that translational and rotational degrees of freedom are independently equilibrated, but at different temperatures. The stationary distribution of a rod in a confining potential $U = \frac{1}{2}k|\bR|^2 + \frac{1}{2}k_\Theta \Theta^2$ is
\begin{equation}
\rhooss \propto e^{-\left(\frac{k}{2}|\bR|^2 + \frac{1}{2M}|\bP|^2
  \right) / \Teff_\bR} e^{-\left(\frac{k_\Theta}{2} \delta\Theta^2 + \frac{1}{2I} L^2 \right) / \Teff_\Theta}\,,
\end{equation}
where $\delta\Theta = \Theta - \Gammaavg / k_\Theta$.
As shown in Fig.~\subfigref{fig:rod-rotation}{b}, the translational
temperature $\Teff_\bR$ is nearly twice as ``hot'' as $\Teff_\Theta$.

The effective equilibrium observed for both the disk and rod can be rationalized by noting that the adiabatic dynamics of rotationally-symmetric objects are statistically reversible under combined time reversal and chirality inversion ($\omega_0 \rightarrow -\omega_0$)~\cite{companionPRE}. 
This hidden symmetry is lost when $C_n$ symmetry is broken, as for the wedge studied in the next section.

\vspace{0.1in}
\noindent\textbf{\textit{The $\Pi$ steering wedge}.}
A wedge, pictured in Fig.~\subfigref{fig:cartoon}{c}, has no $C_n$
symmetry, and the full-system parity symmetry ($\Pi$) is broken by the bath chirality. A wedge thus experiences net forces and torques from the bath, behaving as both a rotational and translational ratchet with average velocities $v_\parallel = M^{-1}\langle \bm{u}(\Theta)\cdot \bP\rangle$ and $v_\perp = M^{-1}\langle \big(\bA\bm{u}(\Theta)\big)\cdot \bP\rangle$, where $\bm{u}(\Theta) = [\cos(\Theta), \sin(\Theta)]^\mathrm{T}$.  In the achiral limit ($\ell_g \rightarrow \infty$) $\Omega$ and $v_\perp$ vanish by symmetry.  In the opposite limit ($\ell_g \rightarrow 0$) the active force $f_0$ acts over a vanishing small distance, and all three ratchet currents vanish.  Consequently, $\Omega$ and $v_\perp$ are maximal at intermediate values of $\ell_g$ which Fig.~\subfigref{fig:wedge}{a} shows to be selected for by the wedge size, $\ell_\text{wedge}$. Note that, while a wedge in an achiral active bath always propels towards its tip, in a chiral bath $v_\parallel$ reverses at $\ell_g \approx \ell_\text{wedge}$, propelling in the opposite direction.

The lack of $C_n$ symmetry of the wedge implies that rotational and translational motion are coupled in $\blambda$ and $\bzeta$.  This is illustrated in Fig.~\subfigref{fig:wedge}{b}, where the average trajectory of the wedge subject to ``pulling'' in the direction $\bm{u}(\Theta)$ or ``pushing'' in the opposite direction produces asymmetric trajectories with different directions of rotation.  In contrast to the disk and rod, this coupling prevents defining effective rotational and translational temperatures for the wedge, even in the adiabatic limit.

\vspace{0.1in}
\noindent\textbf{\textit{The bath's perspective.}}  Decreasing the
object symmetry thus drives its dynamics increasingly far from an effective
equilibrium. This raises the question as to how the broken object
symmetries influence the structure and flows of the bath itself. To
answer that question, we characterize the universal far-field flow
generated by the object in the chiral bath, building on a procedure
developed for achiral
baths~\cite{Baek2018,granekInclusionsBoundariesDisorder2023}. Far from
the object, the bath dynamics is diffusive, with current $\bj_D =
-\mathbf{D}^\mathrm{b} \bm{\nabla} \rho$, where $\rho(\bm{r})$ is the
bath density and $\mathbf{D}^\mathrm{b} = D_\parallel^\mathrm{b}
\id + D_\perp^\mathrm{b} \bA$ is the bath diffusivity.  Near
the object, in contrast, the flux $\bj$ of the active bath involves
higher orientational moments.  Defining $\bm{\delta j} = \bj -
\bj_D$ and applying the steady-state continuity condition
$\bm{\nabla} \cdot \bj = 0$ yields a Poisson equation for the
density
\begin{equation}\label{eq:density-poisson}
    D_\parallel^\mathrm{b} \nabla^2 \rho = \bm{\nabla} \cdot \bm{\delta j}\,,
\end{equation}
where $\bm{\nabla} \cdot \bm{\delta j}$ acts as a localized source
term.  Equation~\eqref{eq:density-poisson} admits a multipole
expansion in powers of the distance $\bm{r}$ to the object that leads to
\begin{equation}\label{eq:multipole-expansion}
    \rho(\bm{r}) = \rho_0 + \frac{1}{2 \pi \gamma D_\parallel} \left\{ \frac{\bm{p} \cdot \bm{r}}{r^2} + \frac{\bm{r} \cdot \mathbf{q} \bm{r}}{2 r^4} \right\} + \mathcal{O}(r^{-3})\,,
\end{equation}
where $\rho_0$ is the bath density far from the object, and the dipole and quadrupole moments are
\begin{align}
    \label{eq:dipole-moment}
    \bm{p} &= -\int d\bm{r}\ \rho(\bm{r}) \bm{\nabla} V(\bm{r})\,,\\
    \label{eq:quadrupole-moment}
    \tilde{\mathbf{q}} &= -2\int d\bm{r}\ \rho(\bm{r}) \bm{r}  \bm{\nabla} V
    + \frac{2 \gamma}{f_0} \mathbf{D}^\mathrm{b}\bm{m}(\bm{r}) \bm{\nabla} V\,.
\end{align}
and $\mathbf{q}$ is the symmetric, traceless part of $\mathbf{\tilde q}$.
Here, $\bm{F}_i = -\bnabla V(\bm{r}_i)$ in
Eq.~\eqref{eq:EOM-newtonian},
and $\bm{m}(\bm{r})$ is the polar order field, quantifying the
orientational order of the bath.  Note that the integrands in
Eqs.~\eqref{eq:dipole-moment}-\eqref{eq:quadrupole-moment} vanish
everywhere except at the object, where $\bnabla V \ne 0$.
Plugging Eq.~\eqref{eq:multipole-expansion} into $\bj = -\mathbf{D}^\mathrm{b}\bnabla \rho$ for the far-field flux yields
\begin{align}\label{eq:J-multipole}
    \begin{split}
        \bj \simeq -\frac{1}{2\pi \gamma D_\parallel} \mathbf{D}^\mathrm{b} \big(r^2 \id - 2\bm{r}\bm{r}\big) \bigg(\frac{\bm{p}}{r^4}
        + \frac{2 \mathbf{q} \bm{r}}{r^6} \bigg) \,.
    \end{split}
\end{align}
In Fig.~\ref{fig:multipole}, the far-field densities and fluxes induced by the rod and wedge in both chiral and achiral active baths shown to agree with our multipole approximation, with chirality of the bath effectively rotating the far-field distribution with respect to the object. While the rod's leading contribution is quadrupolar, the wedge is dominated by the dipole moment, reflecting its increased departure from equilibrium.  Finally, the rotational flux $j_\phi = \bm{r}\times\bj/|\bm{r}|$ of the chiral bath is observed to exhibit a large net circulation near the object. This is related to the torque on the object through Eq.~\eqref{eq:quadrupole-moment}, showing these effects to be two sides of the same coin.

\vspace{0.1in}
\noindent\textbf{\textit{Tailoring transport.}}
Our results show how a passive object in contact with a chiral active bath is imbued with odd transport and ratchet properties determined by
its \textit{shape}, \textit{mass}, and \textit{size}.
First, the object's symmetry determines which active ratchet behaviors and transport couplings are triggered.
Second, we have shown how to build a Langevin description of  a heavy object, predicting effective equilibrium descriptions that nonetheless retain odd signatures of nonequilibrium.
Finally, our simulations show how tuning the object size to the bath gyroradius $\ell_g$ maximizes the odd response coefficients and chiral ratchet motion. An exciting prospect is now to explore the interactions between multiple objects mediated by chiral active bath and the possibility of tailoring their self-assembly.

\vspace{0.1in}
\noindent\textbf{\textit{Acknowledgements.}}
 CH, FG, JT and FvW acknowledge the financial support of the ANR THEMA AAPG2020 grant.

\bibliography{ref}

\onecolumngrid

\newpage
\begin{center}
    {\large \bf End Matter}\\
\end{center}

\begin{figure*}[h]
    \centering
    \includegraphics[width=\textwidth]{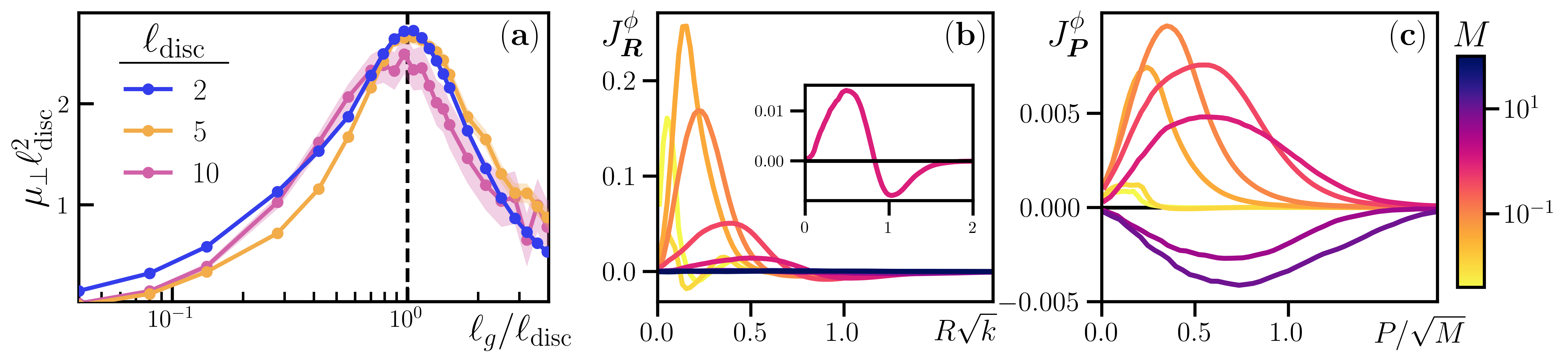}
    \caption{
    \textbf{Disk odd transport} \textbf{(a)} The odd mobility $\mu_\perp$ is maximized when the gyroradius $\ell_g$ of the bath is of the order of the disk diameter $\ell_\mathrm{disk}$. Error bands give 95\% confidence interval from independent replicas of pulling simulations.
    \textbf{(b)} Profiles of the steady-state rotational currents in position space $J_\bR^\phi = \bR \times \bm{J}_\bR^{\text{ss}} / |\bR|$ and \textbf{(c)} momentum space $J_\bP^\phi = \bP \times \bm{J}_\bP^{\text{ss}} / |\bP|$
    corresponding to Fig.~\subfigref{fig:disk-adiabaticity}{c,d} for various values of $M$.
    The inset in (b) shows the $M=1$ profile plotted in Fig.~\subfigref{fig:disk-adiabaticity}{c}, emphasizing counter-rotation.
    \label{fig-EM:disk-end-matter}
    }
\end{figure*}
\twocolumngrid

\noindent\textbf{\textit{Simulation details.}}
Molecular dynamics simulations were carried out using the LAMMPS simulation environment~\cite{lammps} with custom modifications~\cite{github} for the dynamics of the chiral active bath. The bath particles interact with the object through the purely repulsive Weeks-Chandler-Andersen (WCA) potential defined by
\begin{equation}
    V(\bm{r}) = \begin{cases}
          4\epsilon \bigg[ \big(\sigma/\bm{r}\big)^{12} - \big(\sigma/\bm{r}\big)^6 \bigg] + \epsilon & |\bm{r}| < 2^{1/6}\sigma \\
          0 & |\bm{r}| \geq 2^{1/6}\sigma
       \end{cases}
    \end{equation}
with $\sigma = \epsilon = 1$.
The total interaction energy is then $V^\mathrm{int} = \sum_{i=1}^N \sum_{j=1}^{M_\mathrm{object}} V(|\bm{r}_i - \bm{R}_j|)$, where the passive object (in the case of the rod and wedge) is constructed out of $M_\mathrm{object}$ point particles spaced at intervals of $0.1\sigma$. The total force on the object is then $\bm{F} = -\sum_{j=1}^{M_\mathrm{object}} \frac{\partial}{\partial \bR_j} V^\mathrm{int}$.
The passive object dynamics of Eq.~\eqref{eq:EOM-newtonian} were integrated using the velocity Verlet algorithm while the stochastic dynamics of the bath in Eq.~\eqref{eq:cABP-EOM} were treated with a first-order Euler-Maruyama integrator, with a timestep of $\delta t = 0.005$. For the self-propulsion force and bath friction we set $f_0 = \gamma = 1$, so that the $\ell_p$ and $\ell_g$ are always adjusted by changing $D_r$ and $\omega_0$.

The disk was harmonically confined by $U = \frac{1}{2}k|\bR|^2$ with $k=0.08$. Radial profiles of the steady-state fluxes are plotted in Fig.~\ref{fig-EM:disk-end-matter}.
The disk mobility $\bm{\mu} = \bzeta^{-1}$ was measured by pulling with an external force $|\bm{F}^\mathrm{ext}| = 0.1$.
The rod was harmonically confined by $U = \frac{1}{2}k|\bR|^2 + \frac{1}{2}k_\Theta \Theta^2$ with $k=2$ and $k_\Theta=5$, which was sufficient to prevent any net rotation of the rod.
The length of the rod and of the arms of the wedge (set at an angle of $\pi/3$) were fixed at $\ell_\mathrm{rod}=\ell_\mathrm{wedge}=10$. All calculations were performed in a simulation box with periodic boundaries and size $100 \times 100$, except for the multipole calculations where a larger size of $300 \times 300$ was used to sample the far-field effects. In those simulations, Ewald summation was used with a screening length of $\alpha=1/300$ to account for the non-converging contribution of periodic images to the dipole contribution to $\rho(\bm{r})$, which decays as $1/|\bm{r}|$.

\end{document}